\documentclass[aps,pra,twocolumn,groupedaddress,showpacs]{revtex4-1}

\usepackage{graphicx}
\usepackage{amsmath}
\usepackage{textcomp}
\usepackage{mathrsfs}
\usepackage{amsfonts}

\begin{document}

\newcommand{\beq}{\begin{equation}}
\newcommand{\eeq}{\end{equation}}
\newcommand{\barr}{\begin{eqnarray}}
\newcommand{\earr}{\end{eqnarray}}
\newcommand{\bseq}{\begin{subequations}}
\newcommand{\eseq}{\end{subequations}}

\newcommand{\vett}[1]{\mathbf{#1}}
\newcommand{\uvett}[1]{\hat{\vett{#1}}}
\newcommand{\mat}[4]{\left[
\begin{array}{cc}
#1 & #2 \\ #3 & #4 \\
\end{array}
\right]}
\newcommand{\barm}{\bar{m}}
\newcommand{\barv}{\tilde{\nu}}

\title{Analogue gravity by an optical vortex. Resonance enhancement of Hawking radiation.}

\author{Marco Ornigotti$^1$}

\email{marco.ornigotti@uni-rostock.de}

\author{Shimshon Bar-Ad$^2$}
\author{Alexander Szameit$^1$}
\author{Victor Fleurov$^2$}

\affiliation{$^1$Institut f\"ur Physik, Universit\"at Rostock, Albert-Einstein-Stra\ss e 23, 18059 Rostock, Germany}
\affiliation{$^2$ Raymond and Beverly Sackler Faculty of Exact Sciences, School of Physics and Astronomy, Tel-Aviv University, Tel-Aviv 69978, Israel}

\date{\today}

\begin{abstract}
Propagation of coherent light in a Kerr nonlinear medium can be mapped onto a flow of an equivalent fluid. Here we use this mapping to model the conditions in the vicinity of a rotating black hole as a Laguerre-Gauss vortex beam. We describe weak fluctuations of the phase and amplitude of the electric field by wave equations in curved space, with a metric that is similar to the Kerr metric. We find the positions of event horizons and ergoregion boundaries, and the conditions for the onset of superradiance, which are simultaneously the conditions for a resonance in the analogue Hawking radiation. The resonance strongly enhances the otherwise exponentially weak Hawking radiation at certain frequencies, and makes its experimental observation feasible.
 \end{abstract}


\maketitle

\section{Introduction}

Analogue gravity is a research field aimed at creating table-top experimental systems which model processes generally described within the framework of General relativity (GR). This research field essentially originated from the seminal paper by Unruh in 1981 \cite{U81}, where the analog of Hawking radiation \cite{H75,H76} in a transonically-accelerating inviscid barotropic fluid in linear geometry is discussed. In his work, Unruh shows that the accelerating flow in linear geometry creates a background, which mimics curved space with the Schwartzschild metric, and that weak fluctuations with respect to such background are described by the corresponding Klein-Gordon equation (see Ref. \cite{V98} for detailed explanations). More recently, several different physical systems were theoretically proposed, in which the necessary conditions for the onset of a Schwartzschild metric can occur, such as $^3He$ \cite{JV98}, solid state systems \cite{R00}, one dimensional Fermi liquids \cite{G05}, Bose-Einstein condensates (BECs) \cite{BLV03,CFRBF08,RPC09}, superconducting devices \cite{NBRB09} and optical fluids \cite{M08,MCO09,FFBF10,BCP14}, to name a few. Moreover, ``horizon physics" for surface waves in a water channel has also been investigated  \cite{RMMPL08,RMMPL10,WTPUL11,PEPR16,EMPPR16} and, recently, the possibility for a ``magnonic" black hole has been discussed as well \cite{JMPR11,RND17}. Parallel to theoretical proposals, a significant progress in the experimental realization of analogue gravity systems has also been made, like the observation of a white hole horizon in optical fibers \cite{PKRHK08,BCCGORRSF10}, or the realization of a black-hole horizon in BECs by J. Steinhauer and co-workers \cite{LIBGRZS10}, which also reported on the first evidence of Hawking radiation in such a system  \cite{S15}. Moreover, stimulated amplification of Hawking radiation \cite{S14}, in accordance with the predictions of Ref. \cite{CJ99}, has also been reported.

Nearly in all the aforementioned works, however, the background-induced metric is always the same, namely the Schwartzschild metric, which describes the spacetime in the vicinity of an ordinary, non rotating, back hole. In GR, on the other side, there are different metrics that admit black holes as a solution. It would be therefore very interesting to construct analogue models for other types of black hole metrics and to study the effects of these alternative geometries on the process of Hawking radiation.
For example, it would be of particular interest to realize an analogue of a rotating black hole. In this case, the relevant metric would be the Kerr metric \cite{V98}, rather than the standard Schwartzschild metric. Moreover, in such a geometry one would be able to observe not only Hawking radiation, but also superradiance (SR), i.e., the conditions when an incident wave may be amplified by the rotating black hole itself, so that the reflected wave is stronger than the incident one. A vortex in a fluid, in particular, is an exciting possibility for studying the dynamics of fields in the vicinity of rotating black holes. In such a system, the vortex induces a Kerr-type metric \cite{V98} and essentially plays the role of the rotating black hole. In particular the SR effect for the case of vortices in shallow water \cite{WTPUL11}, as well as for BEC \cite{SS05} has been predicted. Very recently, moreover, SR from a vortex in shallow water has also been reported experimentally \cite{TPCRTW16}.

Water waves and atomic systems, however, are not the only media, in which vortices appear. Vortices, in fact, are also known to occur in optics. As shown by the pioneering works of Berry and Nye in 1974 \cite{berry} and Allen and Woerdman in 1992 \cite{woerdman}, optical fields that carry phase singularities, e.g. Laguerre-Gaussian beams, have transverse intensity profiles with all the characteristics of a vortex \cite{OAMbook}. Moreover, it is also well known that coherent light propagation in defocusing nonlinear Kerr media \cite{kerrNote} is analogous to the flow of a fluid, and even a superfluid, by virtue of the so-called hydrodynamic approach to Maxwell's equations. This approach was instrumental for investigating dispersive shock waves \cite{WJF07,EGKKK07,KGYGK08} and tunneling processes \cite{DFSS07}, and its application to the field of analogue gravity was discussed theoretically and experimentally in Refs. \cite{M08,MCO09,FFBF10,EFB12,EBFS13}. The fluctuations in such equivalent photon fluids are predicted to be of the Bogolubov type \cite{BCP14,R12,C13,BF14,LC15}, and recent measurements of their dispersion relation \cite{VRMWCCF15} support this prediction.

Yet, despite a considerable volume of work dealing with the hydrodynamic approach to Maxwell's equations, to the best of our knowledge a comprehensive theory of Hawking radiation from a black hole event horizon in a vortex background has not been studied. The only exception, however, is represented by the works of Marino and co-workers \cite{M08,MCO09}. In these works, the hydrodynamic approach is used to describe the propagation of light in an optical cavity filled by a defocusing medium. In particular, they carefully show how an analogue of a rotating black hole in such a system can be realised by suitably controlling the properties of the vortex state sustained by the cavity. However, the principal attention of these works is concentrated on the study of SR and although the possibility of using this system to study Hawking radiation is envisaged, this phenomenon is not studied in detail.

Hence, we devote this paper to this open question. In particular, we consider Laguerre-Gaussian beam propagating in a defocusing Kerr nonlinear medium, and study the dynamics of fluctuations of the electromagnetic field on such a vortex background, leading to Hawking radiation and SR. We discuss the strong connection between these two phenomena, and show that the conditions for the onset of SR coicide with resonance enhancement of certain frequencies of Hawking radiation.

This paper is organised as follows: in Section \ref{II} we shortly review the hydrodynamic formulation of the nonlinear Schr\"odinger equation in nonlinear optical media, and cast the problem for analysis, namely the field fluctuations in a non-stationary vortex background with a radial flow. In Section \ref{III} we discuss how to obtain the non-vanishing radial flow that is essential for the appearance of an event horizon in the vortex background. Section \ref{IIIbis} is then devoted to analyze the induced "spacetime" geometry, including the positions of event horizons and ergoregions.  Section \ref{IV} is then dedicated to calculation of the Hawking temperature as a function of the background vorticity, and to discuss the spectral density of Hawking radiation, with a particular emphasis on the occurrence of its resonant enhancement, which is essentially due to the background vorticity. Finally, conclusions are drawn in Section \ref{conclusions}. Moreover, the occurrence of SR is investigated in detail in Appendix \ref{AppC}.

\section{Field Fluctuations in a Vortex Background with Radial Flow}
\label{II}

\subsection{Hydrodynamic Formulation of the Nonlinear Schr\"odinger Equation}

The propagation of electromagnetic waves in Kerr nonlinear media can be described, under the paraxial  and the slowly varying envelope  (SVEA) approximations, by the following nonlinear Schr\"odinger equation \cite{B08}:
\beq\label{eq1}
i\frac{\partial A}{\partial z}=-\frac{1}{2\beta_0}\nabla_\perp^2A+g|A|^2A,
\eeq
where the propagation distance $z$ plays the role of time,  $\vett{R}=\{r,\phi\}$, $A\equiv A(\vett{R}, z)$ is the slowly varying amplitude of the electric field propagating in the medium along the $z$ direction, $\nabla_\perp^2$ is the transverse Laplace operator with respect to the variables $x$ and $y$, or $r$ and $\phi$ in polar coordinates, $\beta_0= \omega_0n_0/c = k_0n_0$ is the wave vector of the field in the medium, and $\omega_0$ is the laser frequency.
The parameter $g = 2\beta_0n_2/n_0$ describes the strength of the nonlinear interaction of the laser EM field with the medium, with $n_2$ being the Kerr nonlinear refractive index, i.e., $n(A) = n_0 +  n_2|A|^2$. Applying the Madelung transformation \cite{M27,BEC} $A(\vett{R},z)= f(\vett{R},z) \exp{[-i\varphi(\vett{R},z)]}$,  we obtain the following coupled differential equations:
\bseq\label{eqs2}
\begin{align}
\frac{\partial\rho}{\partial z}+\nabla_{\perp}\cdot\left(\rho\vett{v}\right)&=0,\label{eq2a}\\
\frac{\partial\vett{v}}{\partial z}+\frac{1}{2}\nabla_{\perp}\left(\vett{v}\cdot\vett{v}\right)&=-\frac{1}{\beta_0} \nabla \left[-\frac{1}{2\beta_0}\frac{\nabla_{\perp}^2f}{f}+g\rho\right] \label{eq2b}
\end{align}
\eseq
for the density $\rho(\vett{R},z)= f^2(\vett{R},z)$ and the velocity $\vett{v}= -(1/\beta_0) \nabla \varphi (\vett{R},z)$. The first term on the right hand side of Eq. \eqref{eq2b} is the so-called quantum potential, which accounts for dispersion in the medium. Equations \eqref{eqs2} can be seen as the continuity and Euler equations for a fluid characterized by density $\rho$ and velocity $\vett{v}$. In this form, light dynamics in a Kerr nonlinear medium is similar to the dynamics of a compressible fluid. Usually, the next step would be to consider small fluctuations  $A = \Phi_0 + \psi$ around a $z$-stationary solution $\Phi_0$, i.e.,  the function $\Phi_0$ $z$-dependence is only in the factor $e^{i\mu z}$. However this assumption is not only unnecessary but also undesirable, since creation of a $z$-stationary flow is unrealistic for flow fields that involve radial velocities (radial flow appears only if the beam profile varies with $z$). We therefore assume that the $z$-dependent function $\Phi_0=f_0\exp{(-i\varphi_0)}$, and the functions $\rho_0(\vett{R},z)$ and $\vett{v}_0(\vett{R},z)$, solve Eqs. (\ref{eqs2}) and write the corresponding density and velocity fluctuations in the form $\delta\rho(\vett{R},z)=\rho_0(\vett{R},z)\chi(\vett{R},z)$ and $\delta\vett{v}(\vett{R},z)=-(1/\beta_0)\nabla\xi(\vett{R},z)$. Then, Eqs. \eqref{eqs2} can be linearized and written as follows:
\bseq\label{eqs4}
\begin{align}
\hat{\mathcal{D}}\chi-\frac{1}{\beta_0 \rho_0} \nabla_\perp\left(\rho_0 \nabla_\perp\xi\right)&=0,\label{eq4a}\\
\hat{\mathcal{D}}\xi+ \frac{1}{4\beta_0 \rho_0} \nabla_\perp\left(\rho_0\nabla_\perp\chi\right)- g\rho_0 \chi&=0,\label{eq4b}
\end{align}
\eseq
where $\hat{\mathcal{D}} = \partial_z+\vett{v}_0\cdot\nabla_\perp$. The above set of equations is equivalent to the one obtained for fluctuations on a $z$-stationary background \cite{VF16}. However, in this case, the density $\rho_0$ and velocity $v_0$ are weakly $z$ dependent

\subsection{Fluctuations in a Vortex Background}
If we neglect the quantum potential in Eq. \eqref{eq4b}, solve it with respect to $\chi$ and substitute into Eq. \eqref{eq4a}, we get the Klein-Gordon equation
\beq\label{KG1}
\frac{1}{\sqrt{\det(-g_{\mu\,\nu})}}\partial_{\mu} \left(g^{\mu\nu}\,\sqrt{\det(-g_{\mu\,\nu})}\,\partial_{\nu}\,\xi\right)=0
\eeq
for the phase fluctuation $\xi$ in the curved space determined by the background flow of the $z$-nonstationary solution
$\Phi_0(\vett{R},z)$ \cite{FS12}. The contravariant metric in polar coordinates is then
\beq\label{metric}
g^{\mu\,\nu}= \frac{1}{s} \left(
\begin{array}{cccc}
1 & v_r &\frac{v_{\phi}}{r}&0\\
v_r & v_r^2-s^2 & \frac{v_rv_{\phi}}{r}&0\\
\frac{v_{\phi}}{r} & \frac{v_rv_{\phi}}{r} & \frac{(v_{\phi}^2-s^2)}{r^2}&0\\
0&0&0&-s^2
\end{array}
\right),
\eeq
where we assume that the background flow velocity $\vett{v}_0 = v_r\mathbf{\hat{r}} + v_{\phi}\hat{\boldsymbol{\phi}}$ contains both a rotational and an azimuthal component. The quantity $s$ is the sound velocity of the background flow, defined as follows:
\begin{equation}\label{sound}
\beta_0 s^2=gf_0^2   .
\end{equation}
Although we consider $t$-stationary solutions, a $4 \times 4$ metric is used as a matter of convenience. The fourth coordinate (measured in properly chosen units) is redundant, and can be omitted whenever necessary.

By inverting Eq. \eqref{metric} we find the covariant metric describing the background, namely
\beq\label{gmunu}
g_{\mu\,\nu} =\frac{1}{s}
\left(
\begin{array}{cccc}
s^2-v_0^2 & v_r &r v_{\phi}&0\\
v_r & -1 & 0& 0\\
r v_{\phi} & 0 & -r^2&0\\
0&0&0&-1
\end{array}
\right),
\eeq
where $v_0^2 =v_r^2 + v_{\phi}^2$. In the general case equation (\ref{gmunu}) represents a Kerr-type metric, and therefore delineates two special contours, corresponding to the boundary of the ergoregion and the event horizon of a rotating black hole \cite{general_el}. To let them explicitly appear in the above metric, we first introduce the generalised tortoise coordinates
\begin{eqnarray}\label{transf}
    d\widetilde{z} &=& dz + \frac{v_r}{s^2 - v_r^2}dr, \nonumber \\
    d\widetilde{r} &=& dr, \\
    d\widetilde{\phi}& = &  \frac{v_rv_\phi}{r(s^2 - v_r^2)} dr + d\phi, \nonumber
  \end{eqnarray}
such that the interval $d\sigma^2=g_{\mu\nu}dx^{\mu}dx^{\nu}$ becomes
\begin{eqnarray}\label{interval}
d\sigma^2 &=&  \frac{1}{s} \Big[(s^2 -v_0^2)d\widetilde{z}^2 - \frac{s^2}{s^2 - v_r^2} d\widetilde{r}^2\nonumber\\
&-& r^2d\widetilde{\phi}^2  + 2rv_{\widetilde{\phi}} d\widetilde{z} d\widetilde{\phi}\Big].
\end{eqnarray}
The radius $r_e$ of the ergoregion is then found from the condition $g_{zz}=0$, i.e. $v_0^2(r_e) = s^2(r_e)$, whereas the radius $r_h$ of the event horizon corresponds to the point where $g_{rr}$ diverges, i.e. $v_r^2(r_h) =s^2(r_h)$.

For the case of a background whose $z$-stationary solution $\Phi(\vett{R})$ is a vortex of charge $n$, it follows directly from Eqs. \eqref{eqs2} that $\vett{v}_0=v_{\phi}\hat{\boldsymbol{\phi}} = n/(\beta_0r)\hat{\boldsymbol{\phi}}$, i.e., there is no radial flow, and, therefore, $v_r =0$ \cite{BEC,landau}. Substituting this result into Eq. \eqref{gmunu} we see that the metric for a pure $z$-stationary vortex background contains only one singular point, corresponding to the ergoregion.

The dynamics of fluctuations in a vortex background therefore always admit SR, as the ergoregion, according to Eq. \eqref{gmunu}, is always well-defined. This effect was considered for several model systems \cite{BM03,SS05,RPLW15}. However, the lack of radial flow, i.e., a radial component of the velocity $\textbf{v}_0$ of the vortex, and the consequent absence of an event horizon, does not allow analysis of the effect of the background vorticity on the Hawking process at the event horizon of a rotating black hole. Introducing radial flow requires consideration of a $z$-nonstationary vortex background, as discussed in detail in the following section.

\section{Radial flow on a vortex background}
\label{III}

The absence of a radial component of the velocity flow $\vett{v}_0$ in a $z$-stationary vortex background is essentially due to Eq. \eqref{eq2a}. In fact, for any $z$-stationary solution $\Phi_0(\vett{R})$ of Eq. \eqref{eq1}, Eq. \eqref{eq2a} implies that $\nabla_\perp \cdot(\rho_0\vett{v}_0)=0$. This condition leads immediately to $v_r=0$. The metric for a pure $z$-stationary vortex background contains only one singular point, which corresponds to the ergoregion. In such a metric, therefore, no event horizon appears. Short of introducing source or sink, we now have to look for weakly $z$-dependent solutions $\Phi(\vett{R},z)$ of Eq. \eqref{eq1}.
For the case of an optical beam propagating in a defocusing Kerr medium, we can (at least to the first order in $z$) assume that the solution to Eq. \eqref{eq1} can be sought in the form of an adiabatically slowly varying paraxial vortex beam, e.g. a Laguerre-Gaussian beam. Although this is rigorously true only for the linear case (i.e., $g=0$), we can assume that the effect of the defocusing nonlinearity is only to introduce a nonlinear phase shift that does not drastically affect the form of the solution, at least to the first perturbation order.

With this in mind, let us assume that the density and velocity of the quasi-stationary solution $\Phi(\vett{R},z)$ of Eq. \eqref{eq1} can be written as
\bseq\label{density-b}
\begin{align}
\rho(r,z)&= f_0^2(r,z) =I P(r,z),\label{density}\\
\nonumber\\
\vett{v}(r,z) &=-\frac{1}{\beta_0}\nabla_{\perp}\varphi_0(r,z)=\frac{r}{R(z)}\widehat{\vett{r}}- \frac{n}{\beta_0 r} \widehat{\boldsymbol\phi},\label{velocity}
\end{align}
\eseq
where $I$ is the total intensity of the laser beam and
\begin{equation}
P(r,z)=\frac{2}{\pi |n|!w^2(z)} \left(\frac{2r^2}{w^2(z)}\right)^{|n|} e^{-2r^2/w^2(z)}
\end{equation}
is the normalized intensity profile of a Laguerre-Gaussian beam with
\barr
w^2(z)&=&w_0^2\left[1+\left(\frac{z}{z_R}\right)^2\right],\\
\frac{1}{R(z)}&=&\frac{z}{z^2+z_R^2}
\earr
being its z-dependent width and wavefront curvature, respectively. Moreover, $z_R=\beta_0w_0^2/2$ is the Rayleigh range. As can be seen from Eq. \eqref{velocity}, the radial part of the velocity is related to the wavefront curvature of the beam. Note that at the beam waist, $z=0$, the wavefront is plane and the radial velocity is zero. Rather than positioning the experimental apparatus far away from the beam waist, where $R(z) \approx z$, we choose to position a defocusing lens with focal length $-f$ at the waist, a short distance $z$ before the input plane of the nonlinear medium, so that the phase front of the beam is no longer planar, resulting in a radial velocity that monotonically increases (from zero on-axis) along the radial coordinate. Following standard Gaussian optics \cite{PP13} (see Appendix \ref{radial}),  the intensity and velocity profiles of the field at the input plane of the nonlinear medium are given, in the limit of small $z$, by\\
\bseq\label{eqs11}
\begin{align}
P(r,z)&=P_0(r)\left[1+P_1(r)z\right],\label{eq11a}\\
\nonumber\\
\vett{v}(r,z)&=\frac{r}{f}\mathbf{\hat{r}}-\frac{n}{\beta_0r} \hat{\boldsymbol{\phi}}+O(z),\label{eq11b}
\end{align}
\eseq
where
\barr
P_0(r)&=&\frac{2}{\pi |n|! w_0^2}\left(\frac{2r^2}{w_0^2}\right)^{|n|}e^{-2r^2/w_0^2},\\
P_1(r)&=&\frac{2}{f w_0^2}\left[2r^2-(|n|+1)w_0^2\right].\label{P1}
\earr
It is not difficult to show that these density and velocity fields satisfy both the continuity and the Euler equations, up to the order $O(z/f)$. Crucially, the background velocity $\vett{v}_0$ now has a radial component: $v_r=r/f$.

As mentioned before, the defocusing nonlinearity adds a nonlinear phase, which essentially acts as a nonlinear defocusing lens \cite{B08}. The above equations can be corrected to account for this effect by simply setting $1/f= 1/f_L + 1/f_{NL}$. In this case, $f_L$ accounts for the linear radial flow induced by the lens at the beam waist, while $f_{NL}$ is the focal length of the equivalent defocusing lens generated by the nonlinear defocusing. Ultimately, $f_{NL}$ is related to the nonlinear length of the Kerr-medium \cite{B08} and accounts for a nonlinear correction to the radial flow.

This very simple experimental configuration allows us to fully explore the effects of vorticity, not only in terms of SR scattering from the ergoregion, as in Refs. \cite{BM03,SS05,RPLW15}, but rather in terms of the dynamics of fluctuations in the vicinity of the event horizon of the vortex background. In what follows, we will use this model to study the effect of the background vorticity on both Hawking radiation and SR.
\section{Induced Spacetime Geometry}\label{IIIbis}
\subsection{Event horizon}

\begin{figure}[tbp!]
\begin{center}
\includegraphics[width=1\columnwidth]{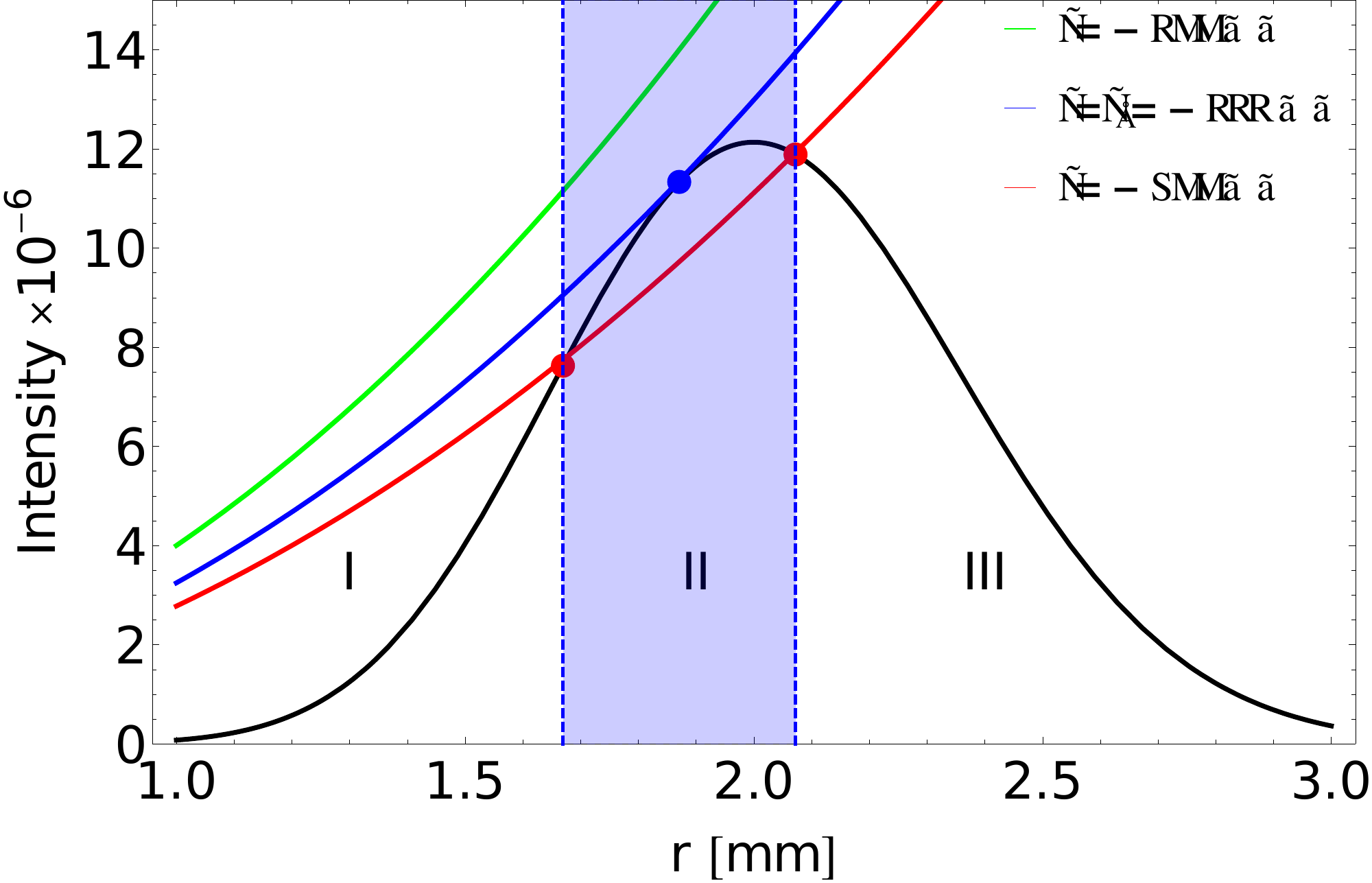}
\end{center}
\caption{(Color online) Graphical solution of equation (\ref{horizon}). The bell shaped curve (in black) represents the Laguerre-Gauss profile of the laser beam, while the 3 coloured curves (green, blue and red) represent the squared radial velocity $v_r^2(r)$ for three different values of the focal length $f$, corresponding to $f<f_c$, $f=f_c$ and $f>f_c$, respectively. For $f=-500$m (upper, green curve) there are no solutions. For $f_c = -555$ mm (middle, blue curve) there is one solution. For $f = -600$ mm (lower, red curve) there are two solutions. In the latter case, two event horizons appear, thus introducing a subsonic (region II, shaded in blue in the figure) and a supersonic (regions I and III) region for the flow. The radial intensity profile of the vortex, with vorticity $n=8$, corresponds to $I =2$ W, $ g= 5.5 \cdot 10^{-4}$ m/W, $w_0 = 1$ mm, and $\beta_0 = (2\pi/7.80) \cdot 10^7$ m$^{-1}$. These parameters allow a broad range of frequencies to satisfy the requirement $L^{-1} < \nu <l_n^{-1}$, where $L$ is the length of propagation in the nonlinear medium, and $l_n$ is the nonlinearity length defined in equation (\ref{eq36}).} \label{horizons1.fig}
\end{figure}

Our first step is to find the positions of the event horizon and ergoregion and to explore the global geometry described by the vortex background. According to Section \ref{II}, the position of the event horizon is determined by the equation $s^2(r)=v_r^2(r)$, namely
\begin{equation}\label{horizon}
\frac{gI}{\beta_0 } P(r) = \frac{ r^2}{f^2}.
\end{equation}
A graphical solution of this equation is shown in Fig. \ref{horizons1.fig}. Depending on the values of the parameters, it may have no solution (upper, green curve), one solution (middle, blue curve) or two solutions (lower, red curve). In the case of a single solution the relation
\begin{equation}\label{horizon-d}
\frac{gI}{\beta_0 } \frac{dP(r)}{dr} = \frac{2r}{f^2},
\end{equation}
must also hold, and thus the solution in this case is
\beq
r_c= \sqrt{\frac{n - 1}{2}} w_0.
\eeq
Obviously, Eqs. \eqref{horizon} and \eqref{horizon-d} admit no solution for $n = 0$, while for $n = 1$ we obtain $r_c=0$. In general, however, there can only be one nonzero solution, depending on the parameters (e.g., the focal length $f$).
Two nonzero solutions appear only when $n > 1$ and $|f |<|f_c|$, where the critical value $f_c/w_0$ depends on the vorticity $n$. For sufficiently small $\delta f = f - f_c$, the two solutions are slightly below and slightly above $r_c$:
$$
r_{h\pm} =r_c  \pm w_0 \sqrt{\frac{\delta f }{2f_c}}.
$$
For larger $\delta f$, the outer horizon $r_{h+}$ falls outside the maximum of the Laguerre-Gauss beam profile (see for example the red curve in Fig. \ref{horizons1.fig}).

\subsection{Ergoregion}

\begin{figure}[tbp!]
\begin{center}
\includegraphics[width=1\columnwidth]{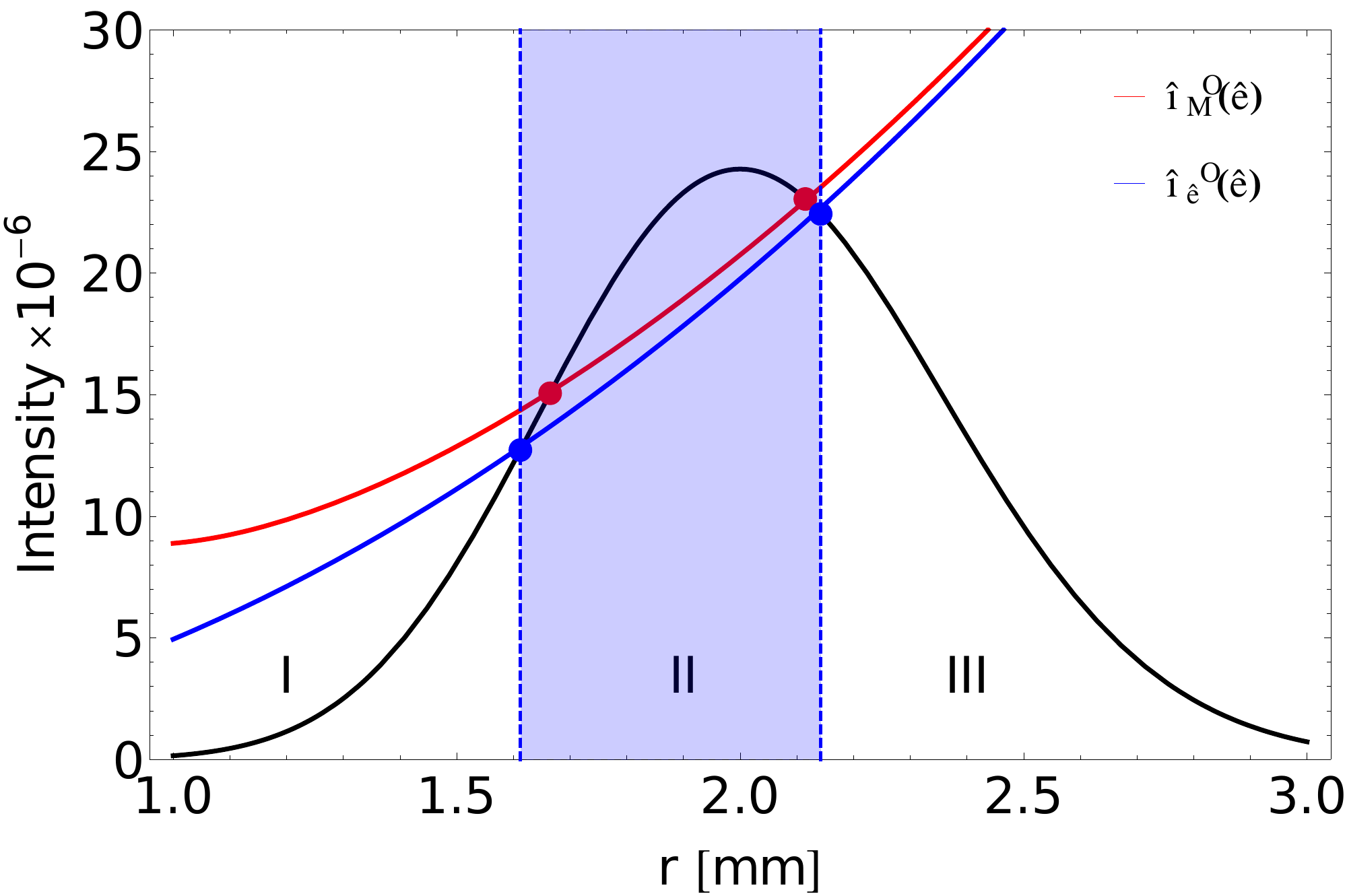}
\end{center}
\caption{(Color online) Graphical solution of equation (\ref{ergoregion}). The bell shaped curve (in black) represents the Laguerre-Gauss profile of the laser beam. The red (upper) curve represents the squared total velocity $v_0^2(r)$ and defines the position of the inner ($r_{e-}$) and outer ($r_{e+}$) ergoregions. The blue (lower) curve corresponds to the squared radial velocity $v_r^2(r)$ and defines the position of the inner ($r_{h-}$) and outer  ($r_{h+}$) horizons.There are two ergoregions, one lying between $r_{h-}$ and $r_{e-}$, and the other lying between $r_{e+}$ and $r_{h+}$. The parameters used here are the same as in Fig. \ref{horizons1.fig}, except for $\beta_0 = (\pi/7.80) \cdot 10^7$ m$^{-1}$ and $f = -450$ mm. The flow in regions I and III is supersonic, whereas in region II it is subsonic.
} \label{ergoregion.fig}
\end{figure}

The other singular point appearing in the metric $g_{\mu\nu}$ given by Eq. \eqref{interval} gives the position of the ergoregion, i.e., the value $r=r_e$ where the total velocity of the fluid equals the background sound velocity, namely $s^2 (r)= v_0^2(r)$. For the case of an optical vortex beam propagating in a nonlinear medium, we get
\begin{equation}\label{ergoregion}
\frac{gI}{\beta_0 } P(r) = \frac{ r^2}{f^2} + \frac{n^2}{\beta^2_0 r^2}.
\end{equation}
A typical graphical solution is shown in Fig. \ref{ergoregion.fig} (lower, blue line), together with the corresponding solution of Eq. (\ref{horizon}) (upper, red line). As it can be seen from Fig. \ref{ergoregion.fig}, we obtain two ergoregions: the outer ergoregion $r_{e+}$, which corresponds to the outer horizon $r_{h+}$ (close to the border between regions II and III in Fig. \ref{ergoregion.fig}, and the inner ergoregion $r_{e-}$, which corresponds to the inner horizon $r_{h-}$ (close to the border between regions I and II in Fig. \ref{ergoregion.fig}). In both cases, moreover, both ergoregions are inside the subsonic region (shaded blue area, in Fig. \ref{ergoregion.fig}).
\begin{figure}[tbp!]
\begin{center}
\includegraphics[width=1\columnwidth]{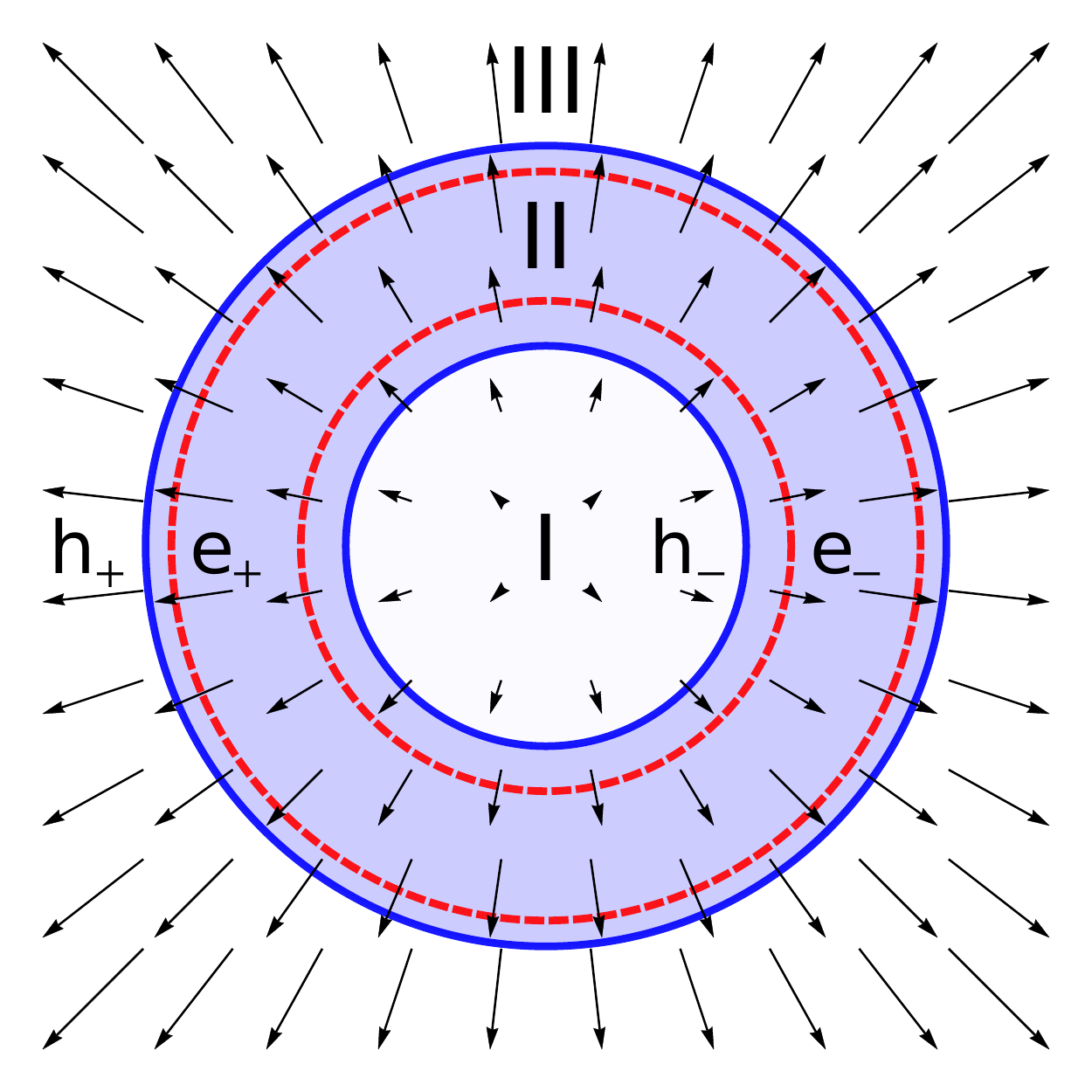}
\end{center}
\caption{(Color online) A schematic depiction of the flow structure of a vortex. Two solid blue circles represent the outer ($h+$) and inner ($h-$) event horizons, separating the supersonic regions I and III from the subsonic region II (shaded blue area). Two dashed circles show the borders of the outer ($e+$) and inner ($e-$) ergoregions. The arrows show the radial component of the outgoing flow.
} \label{flowscheme.fig}
\end{figure}

\subsection{Vortex Geometry}

The resulting 2D geometry of the background, including the positions of the ergoregions and the event horizons, is depicted in Fig. \ref{flowscheme.fig}. The inner ($h_-$) and outer  ($h_+$) horizons are depicted by solid circles, separating three regions: I and III are supersonic regions,  while II is subsonic and sandwiched between them (shaded blue area in Fig. \ref{flowscheme.fig}). It is important to notice, that for the case of a black horizon the radial component of the flow is directed towards the horizon in the subsonic region, crosses the horizon and enters the supersonic region. In a white horizon, on the other hand, the direction of the radial flow is from the supersonic to subsonic region. The direction of the azimuthal component of the velocity plays no role in this case. Therefore, in the case of the outgoing flow considered here, the outer horizon, $h_+$, is black, whereas the inner horizon, $h_-$, is white. In the case of ingoing flow the roles of $h_+$ and $h_-$ are reversed. The two ergoregions $e_+$ and $e_-$ (dashed circles in Fig. \ref{flowscheme.fig}) are located inside the subsonic region.

This constitutes a significant difference with respect to the Kerr or Kerr-Newman geometry typical of rotating black holes. Although the latter also has outer and inner horizons, the arrangement of areas of sub- and super-luminal escape velocities is the opposite of the one shown here, the ergoregions are positioned differently, and there is no turnaround radius in the optical system  \cite{general_el}. This geometry, moreover, differs from the one presented in Refs. \cite{M08,MCO09}, where only one event horizon is considered and the radial flow is directed towards the center of the vortex, instead of away from it, as in our case.

\section{Hawking radiation with vortex background}\label{IV}
Now we are in a position to analyze the properties of fluctuations near the event horizon. Our approach follows essentially the one used in Refs. \cite{FS12,VF16}, with some modifications in order to take into account the $z$-dependence of the beam profile and the curvature of the horizon. To begin with, let us introduce the new variables $x_\pm = r - r_{h\pm}$, such that $\partial_r=\partial_x$. Here and below, if not specified otherwise we omit the $\pm$ sign, for the sake of clarity. With this definition, the sound velocity and the radial velocity in the vicinity of the horizons can be written as $s_{h}^2(x) = s_{h}^2(1 - \alpha_{s} x)$ and $v_r(x)=s_{h}(1 + \alpha_{r} x)$, respectively. Substituting this into Eqs. (\ref{eqs2}), the following conditions must hold in the leading order in z:
\begin{eqnarray}
\frac{1}{\rho}\frac{\partial\rho}{\partial z} &=& - s_h\left(\alpha_r - \alpha_s +\frac{1}{r_h}\right),\label{nonadiabatic-3}\\
\frac{\partial v}{\partial z} &=& - s_h^2(\alpha_r - \alpha_s),\label{nonadiabatic-4}
\end{eqnarray}
where $s_h$ is the sound velocity at the horizon and $r_h$ is the position of the horizon. The first condition follows from the continuity equation [Eq. \eqref{eq2a}], while the second from the Euler equation [Eq. \eqref{eq2b}]. In the case of the Laguerre-Gaussian beam, we have $s_{h}= r_h/f$, $\alpha_r = 1/r_h$ and
\beq
\alpha_s= -\frac{2|n|}{r_h} + \frac{4r_h}{w_0^2}.
\eeq
Then using Eqs. \eqref{nonadiabatic-3} and (\ref{nonadiabatic-4}) for the case of a Laguerre-Gauss beam gives
\begin{equation}\label{nonadiabatic-1}
s_h(\alpha_r -\alpha_s + 1/r_h) = - P_1(r=r_h).
\end{equation}
The validity of this condition can be directly verified by substituting the above definitions into Eq. (\ref{P1}).

The calculations carried out below assume the adiabatic approximation with respect to the weak $z$-dependence of the background density and velocity. This means that their derivatives with respect to $z$ are discarded, except for nonadiabatic corrections (\ref{nonadiabatic-3}) and (\ref{nonadiabatic-4}). We also take the curvature $1/r_h$ of the event horizon into account. The $z$-nonstationarity and curved horizon are essential corrections, and constitute important differences with respect to the analysis carried out in Refs. \onlinecite{FS12,VF16}. Another assumption is that the position of the black horizon doesn't vary with $z$. Varying Eq. (\ref{horizon}) with respect to $z$ and using Eq. (\ref{P1}) we can understand that this assumption works well if the black horizon lies in the vicinity of the optimal radius
\begin{equation}\label{horizon-f}
r_{opt} =\sqrt{\frac{|n|+1}{2}}w_0
\end{equation}
then $\alpha_s =2/r_h$. For the parameters used above we can estimate the corresponding vorticity as $n_{opt} \approx 22$.

The starting point of our analysis are then Eqs. \eqref{eqs4}, which we now expand with respect to the small parameters $\alpha_{s,r} x \ll 1$ and $x/r_h \ll 1$. Moreover, we take the Fourier transform of the field fluctuations $\chi(x,\phi,z)$ and $\xi(x,\phi,z)$, namely
\beq
\chi(x,\phi,z) = \sum_m \int  d\nu \int dk \chi_{k,m,\nu} e^{i(\nu z - m\phi -k z)},
\eeq
thus obtaining the following set of coupled equations for the Fourier components of the field fluctuations
\beq\label{eqs4-1}
\left( \begin{array}{cc}
 \mathcal{A}_m(k) & \mathcal{B}_m(k)\\
 & \\
\mathcal{C}_m(k)& \mathcal{A}_m(k)
\end{array}\right)\left(
\begin{array}{c}
\chi_{k,m,\nu}\\
\\
\xi_{k,m,\nu}
\end{array}\right)=0,
\eeq
where
\beq\label{nu}
\tilde\nu_m=\frac{1}{s_h}\left(\nu-\frac{mn}{\beta_0r_h^2}\right),
\eeq
is the normalised vortex-corrected frequency and the matrix elements $\mathcal{A}_m(k)$, $\mathcal{B}_m(k)$ and $\mathcal{C}_m(k)$ are given by
\bseq
\begin{align}
\mathcal{A}_m(k)&=i (\widetilde{\nu}_m -k) -i \alpha_r\partial_k k,\\
\mathcal{B}_m(k)&=\frac{1}{\beta_0s_h } \left[( - \alpha_s + 1/r_h) ik  + k^2 + \frac{m^2}{ r_h^2}\right],\\
\mathcal{C}_m(k)&=-\frac{1}{4}\mathcal{B}_m(k)- s_h\beta_0(1 + i \alpha_s \partial_k).
\end{align}
\eseq
 Following the procedure detailed in Appendix \ref{nonadiabatic}, the solution of Eq. (\ref{eqs4-1}) can be written as
\beq\label{solution}
\chi(x,m,z)= e^{i(- m\phi+\nu z)}F(\tilde\nu,x),
\eeq
where
\beq\label{integral35}
F(\tilde\nu,x)=\int_C\, dk\, k^{\gamma_1} \left(k - \frac{2 \widetilde{\nu}}{3}\right)^{\gamma_2} e^{\Lambda_0(k,\widetilde{\nu})-ikx},
\eeq
with the definitions
\bseq\label{eq36}
\begin{align}
\gamma_1 &= \frac {i\widetilde{\nu}_m }{2\alpha_r} - \frac {i {m}^{2}}{{2 r_h}^{2}\widetilde{\nu}_m\alpha_r}
+ O(l_n^2/r_h^2),\label{gamma1}\\
\gamma_2 &= \Big[ \frac {\alpha_s -\alpha_r }{\varrho}  - \frac {1 }{r_h \varrho} - \frac {i\widetilde{\nu}_m }{2 \alpha_r} + \frac {2\,i \widetilde{\nu}_m}{\varrho}\nonumber\\
& + {\frac {i {m}^{2}
}{2r_h^{2}\widetilde{\nu}_m\alpha_r}}\Big]+ O(l_n^2/r_h^2), \label{gamma2}\\
\Lambda_{m,\nu}(k) &= \frac{k^3l_n^2}{\varrho}\left[\frac {i }{6}  + \frac {i}{ 2kr_h}  + \frac {i\widetilde{\nu} \alpha_r}{2 k\varrho} + \frac{\alpha_s}{2k}\right.\nonumber\\
&\left.+ O(1/(kr_h)^2)\right]. \label{Lambda}
\end{align}
\eseq

In the expressions above, $l^2_n = 1/(\beta_0g\rho)$ ($l_n \approx 10^{-5}m$ for the above parameters) is the so-called nonlinearity length, akin to the healing length in BEC \cite{FS12}, and $\varrho=2\alpha_r+\alpha_s$. At $r_{opt}$ we get $\varrho = 4/r_{h+}$. Only the leading terms are retained in Eqs. \eqref{eq36}. The full expressions are given in Appendix \ref{nonadiabatic}. The convergence of the above integral is controlled by the cubic term in the exponential. As a result, one can find four independent contours of integration, corresponding to four independent solutions of Eq. \eqref{eqs4-1}. The integral in Eq. \eqref{integral35} can be solved using the steepest descent technique. The equation for the saddle points then reads
\barr\label{saddle-1}
&&\left[(\widetilde{\nu}s_h -kv_r(x))^2  -  ik( \alpha_r - \alpha_s + 1/r_h) - \frac{m^2}{ r_h^2}\right] \nonumber\\
&=&\frac{l_n^2}{2} \left[( - \alpha_s + 1/r_h) ik  + k^2 + \frac{m^2}{ r_h^2}\right]^2  + k^2s(x)^2.
\earr

\begin{figure}[!t]
\begin{center}
\includegraphics[width=1\columnwidth]{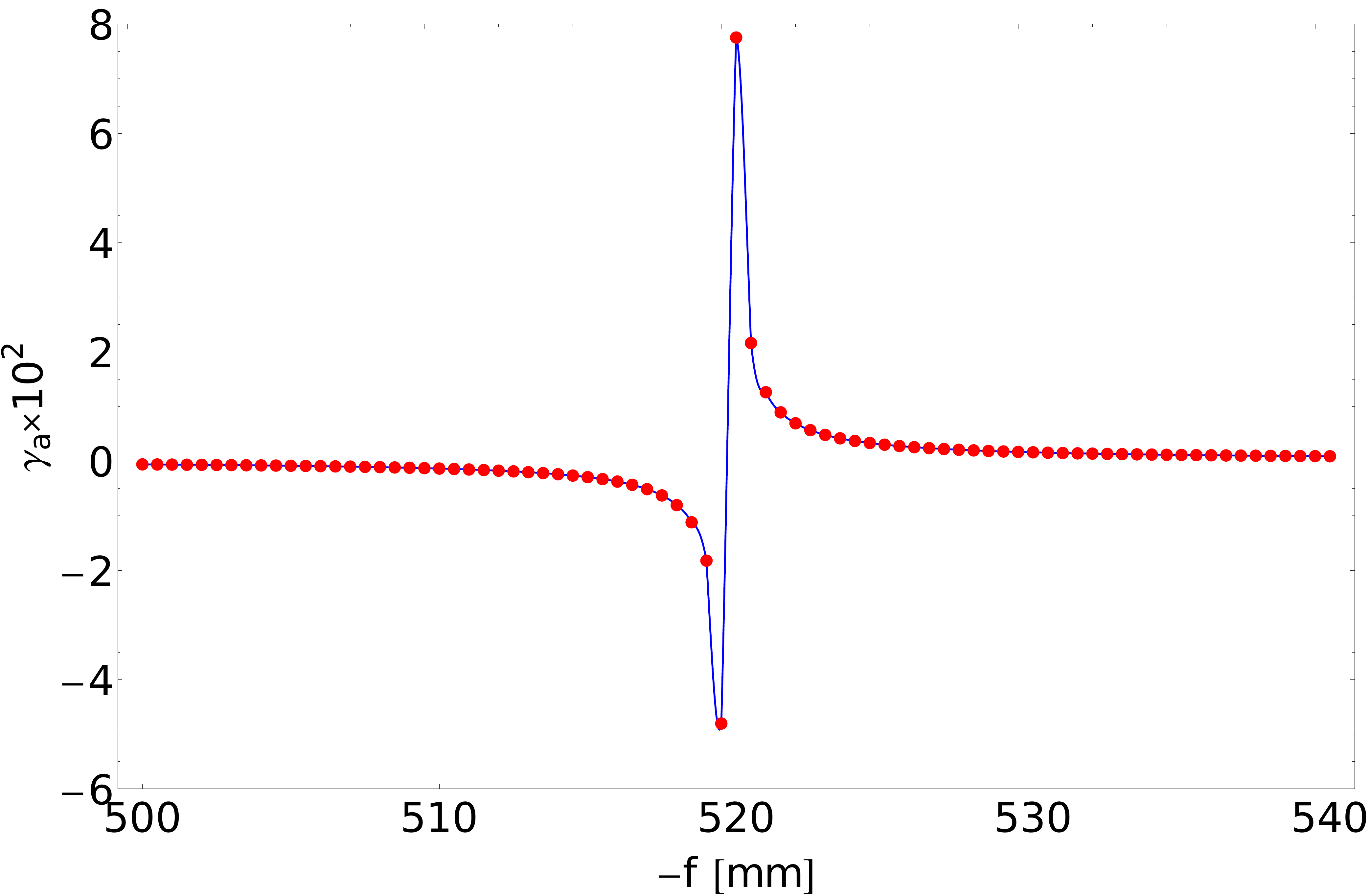}
\end{center}
\caption{(Color online) $\gamma_a$ as a function of the focal length $f$ for the outer horizon. The red dots correspond to the actual numerical value of $\gamma_a$ , while the blue solid line is a spline interpolation. As can be seen, in the vicinity of the critical focal length $f_c\simeq 520 mm$, $\gamma_a$ shows a typical resonant behavior. Note, that while Eq. \eqref{gammaa} contains an actual divergence for $f=f_c$, in the interpolation shown in this picture, this does not appear, as it is instead replaced by a resonance. The divergence, in fact, is an artifact of the approximated analysis carried out to obtain Eq. \eqref{gammaa} and it is not present if the full solution is taken into account. The radial intensity profile of the vortex, with vorticity $n=6$, corresponds to $I =2$W, $ g= 5.5 \cdot 10^{-4}$m/W, $w_0 = 1$ mm, and $\beta_0 = (2\pi/7.80) \cdot 10^7$ m$^{-1}$.} \label{gamma.fig}
\end{figure}

The first two saddle points can be obtained in the limit of small $k$, when the terms $O(l_n^2)$ in Eq. (\ref{saddle-1}) can be neglected. This results in one singular and one regular solution. The singular solution is
\beq
k_s = \frac{2\widetilde{\nu}_m s_h -  i( \alpha_r - \alpha_s  + 1/r_h)}{x\varrho} \propto \frac{1}{x},
\eeq
and corresponds to $\chi_s = x^{\gamma -1}$, where, to the leading order,
\beq\label{gamma_text}
\gamma= -\gamma_1 - \gamma_2= \gamma_a +\gamma_0 + O(l_n^2),
\eeq
with
\bseq
\begin{align}
\gamma_a &= \frac{\alpha_r - \alpha_s}{\varrho} + \frac{1}{r_h\varrho} = \frac{(|n|-1)w_0^2 - 2 r_h^2}{(-2 r_h^2+w^2 (|n|+1)} ,\\
\gamma_0 &=- \frac {2\,i \widetilde{\nu}_m}{\varrho}.
\end{align}
\eseq
The second saddle point is given by
$$
k_r = \frac{ \widetilde{\nu}_m^2s_h^2 - m^2/r_h^2}{2\widetilde{\nu}_ms_h^2  + i(\alpha_r -\alpha_s + 1/r_h)},
$$
and corresponds to the regular solution $\chi \propto e^{-ik_rx}$. Together with the regular and singular solutions displayed above, there are two more solutions, corresponding to evanescent states in the subsonic region II in Fig. \ref{flowscheme.fig}. These solutions, however, become propagating in the supersonic regions I and III. To find them we have to consider the limit of large $k$, $kl_n \gg 1 $, at $x> \alpha l_n^2$. Then, Eq. (\ref{saddle-1}) becomes
\begin{equation}\label{saddle-2}
\frac{l_n^2}{2} k^3 - k\varrho x = 0.
\end{equation}
This equation admits two solutions, namely
\beq
k_{e1,2}= \pm \sqrt{2\varrho x/l_n^2},
\eeq
which correspond to the functions
\beq
\chi_{e1,2} \propto  \exp{\left(\pm i \frac{\sqrt{2\varrho}}{3 l_n}x^{3/2}\right)}.
\eeq
Finally, we carry out the transformation given by Eq. (\ref{transf}) and use the relation between the functions $\chi$ and $\xi$ to obtain the following triads of incoming waves
\bseq\label{incoming}
\begin{align}
      \xi_{r1}(x) &= | x|^{-\gamma_0/2} e^{i\nu z - i m \phi}, & x < 0, \\
      \xi_{r2}(x) &=  x^{-\gamma_0/2} e^{i \nu  z - i m \phi},&  x> 0 , \\
      \xi_{e1}(x) &= \sqrt{\frac{4l_n\widetilde{\nu}_m}{(\varrho\bar x)^{3/2}}} \tilde x^{\gamma_0/2} e^{i\nu z - i m\phi  + i \frac{\sqrt{2\varrho}}{3 l_n}x^{3/2}},
\end{align}
\eseq
and outgoing waves
\bseq\label{outgoing}
\begin{align}
      \xi_{s1}(x) &= | x|^{\gamma_a + \gamma_0/2} e^{i \nu z - i m \phi},& x < 0, \\
      \xi_{s2}(x) &=  x^{\gamma_a + \gamma_0/2} e^{i \nu z - i m \phi},&  x> 0  \\
      \xi_{e1}(x) &= \sqrt{\frac{4l_n\widetilde{\nu}_m}{(\varrho\bar x)^{3/2}}} \tilde x^{\gamma_0/2} e^{i \nu z -i m \phi - i \frac{\sqrt{2\varrho}}{3 l_n}x^{3/2}}.
\end{align}
\eseq
Note that the eigenfunctions $\xi_r$ and $\xi_s$ are propagating in both the subsonic (II) and supersonic (I and III) regions, whereas the eigenfunctions $\xi_{1e}$ and $\xi_{2e}$ are propagating only in the supersonic regions. A detailed discussion of interrelation between these solutions is presented in Refs. \cite{FS12,VF16} for the linear $z$-stationary flow background, and the analysis of the scattering problem outlined there can be fully applied to the present case. Although the eigenfunctions presented here are formally similar to those of Refs. \cite{FS12,VF16} (see also Refs. \cite{C98,CJ96,CPF14,BP12} for an analysis related to the Schwarzschild black hole), there are two important differences. First, the singular eigenfunctions $\xi_{s1,s2}$ acquire now an extra factor $|x|^{\gamma_a}$, where
\begin{equation}\label{gammaa}
\gamma_a = \frac{2(r_m^2- r_h^2) -  w_0^2}{2(r_m^2 - r_h^2) +w_0^2 },
\end{equation}
$r_m^2= |n|w_0^2/2$ being the maximum of the Laguerre-Gaussian beam. The quantity $\gamma_a $ may be either positive or negative, depending on the parameters, which leads either to an increasing density of fluctuations when approaching the horizon if $\gamma_a<0$, or to a suppression of the fluctuation density near the horizon if $\gamma_a > 0$. One also must not forget that this description holds for $|x| > l_n$. Figure \ref{gamma.fig} shows $\gamma_a$ for the outer black horizon $h+$ as a function of the focal length $f$ of the diverging lens (i.e., as a function of the radial flow). For the outer horizon, $r_m < r_{h+}$ and therefore the numerator in Eq. \eqref{gammaa} is always negative, whereas the denominator can be zero and change its sign at $r_{opt}$, Eq. (\ref{horizon-f}). As a result, there exists a critical value of the focal length $f$ (corresponding to a critical value of the radial flow) where $\gamma_a$ has a vertical asymptote (i.e., it diverges and also changes its sign). For negative $\gamma_a$ in this region, the fluctuation density becomes strongly skewed towards the outer event horizon. For the inner event horizon, on the other hand,  $r_m > r_{h-}$ and it is the numerator that goes to zero when $\gamma_a$ changes its sign. Therefore, no divergence is observed in this case.

The second important distinction is connected with the fact that expression \eqref{nu} for the frequency $\widetilde{\nu}$ contains the term reflecting the vorticity $m$ of the mode as well as the vorticity $n$ of the background. A similar problem in the GR context for the Kerr-Newman black hole is discussed in Refs. \cite{H75,C95,U10,ANOP10,MS06}. In particular $\widetilde{\nu}$ can become negative, which implies SR. In order to get the spectrum of Hawking radiation we have to carry out the transformation given by Eq. (\ref{transf}) and solve the scattering problem, in a similar manner as explained in detail in Ref. \cite{VF16}. Following the procedure highlighted in Appendix \ref{norma}, we finally get
\begin{equation}\label{spectraldensity}
N(\nu) = \left[e^{\pi\operatorname{Im}\{\gamma\}}-1\right]^{-1},
\end{equation}
where
\beq
\operatorname{Im}\{\gamma\} = \frac{2\tilde{\nu}}{s_{h}\varrho}=\frac{2}{s_{h}\varrho}\left(\nu - \frac{nm}{\beta_0r_h^2}\right).
\eeq
The corresponding Hawking temperature is then given by
\beq
T_H(\nu) = \frac{\hbar s_h\varrho}{2\pi k_B},
\eeq
where $k_B$ is the Boltzmann constant.

In full analogy to the situation discussed Ref. \cite{H75} (see Eq. (3.2) of that paper) the distribution (\ref{spectraldensity}) reflects the spectral density of the radiation in the immediate vicinity of the horizon. However, in order be registered far from the horizon it must overcome the barrier (\ref{barrier}). Therefore the relevant quantity is
\begin{equation}\label{spectraldensity-a}
J = {\cal T} N(\widetilde{\nu}) = \frac{|T|^2}{\nu}\left(\nu
- \frac{m n}{\beta_0 r_h^2}\right) N(\widetilde{\nu})
\end{equation}
It describes the spectral density of the Hawking radiation reaching the distant observer. This formula emphasizes the connection between the SR and Hawking radiation. It means that the SR can be treated as stimulated emission whereas the Hawking radiation is the corresponding spontaneous radiation.(see Eq. (4.110) in Ref. \onlinecite{PT00}).

\subsection{Resonant enhancement of Hawking radiation}
The situation when $\widetilde{\nu}_m$ is close to zero, i.e., when $\nu \sim nm/(\beta_0r_h^2)$,  is of a special interest, since we expect the radiation to be strongly enhanced in this spectral region. In this case, by Taylor expanding Eq. \eqref{spectraldensity-a} around $\tilde{\nu}_m=0$ and considering only the leading order terms, we have
\beq\label{radiation}
J(\nu)=\frac{ s_h \varrho}{2\pi\nu}|T|^2.
\eeq
The distribution $N(\nu)$ becomes negative at $\widetilde{\nu} < 0$ which indicates an instability due to the onset of SR, when any small seed fluctuation becomes amplified (see, e.g. Ref. \cite{S73,H75,U74,C95,PT00}). Simultaneously the transfer factor ${\cal T}$ changes its sign so that the quantity (\ref{spectraldensity-a}) remains positive. Our estimates in Section (\ref{SR}) allow us to conclude that the transfer coefficient $|T|^2$ is close to one.

The condition $\widetilde{\nu} =0 $ can be then rewritten in the form
\begin{equation}\label{resonance}
 m \lambda_\nu = \tau_\phi,
\end{equation}
where $\lambda_{\nu}=2\pi/\nu$ and $\tau_{\phi}=2\pi r_h/v_{\phi}$, and $v_{\phi}$ is the azimuthal component of the velocity flow. The above expression can be interpreted as a typical resonance condition, which happens when an integer number $m$ of wavelengths $\lambda_\nu$  coincides with the propagation distance $\tau_{\phi}$ necessary for one full rotation of the vortex. This condition is also related to the SR, since it corresponds to the total reflection [see Eq. (\ref{SR-a})]. This resonance condition can strongly enhance the otherwise exponentially weak Hawking radiation at certain frequencies, and makes its experimental observation feasible.

To understand why this is true, one should realise that in order to establish quasi-stationary conditions, and avoid a strong $z$-dependence of the position of the horizon and the physical parameters there, the cell containing the nonlinear medium should be significantly shorter than the local length $f$. This sets a lower limit to the frequency of Hawking radiation that such a cell can emit, namely $\lambda_c = 2\pi/\nu_c < f$. In the absence of vorticity, $n = 0$ when $\varrho \approx 4/r_h$, the spectral weight of the emitted frequency components is exponentially small:
\beq\label{spectraldensity-b}
J = |T|^2 N(\nu_c) \propto e^{-\frac{\nu_c}{\nu_H}} =  e^{-\frac{\pi^2 f}{\lambda_c}}  ,
\eeq
where
\beq
\nu_H = \frac{2\pi}{\lambda_H} = \frac{s_h \varrho}{2\pi  } \approx \frac{2}{\pi f}   .
\eeq
This situation may drastically change in the case of a vortex with a sufficiently high vorticity $n$, since near the critical frequency,
\beq
\left|\nu - \frac{nm}{\beta_0 r_h^2}\right| \sim \frac{1}{\lambda_H},
\eeq
the radiation intensity would not contain the exponentially small factor anymore. One can readily estimate that in order to obtain a resonance at $\lambda_r = 2\pi/\nu_r \approx 10$ cm  (which is a typical propagation length for realistic experimental parameters), the condition
\begin{equation}\label{condition}
\lambda_\nu \approx \frac{2\pi\beta_0 r_h^2}{nm} \approx \frac{\pi\beta_0 w_0^2(n +1)}{nm}
\end{equation}
must be satisfied. Using Eq. (\ref{horizon-f}) the vorticity $n$ drops out at large $n$ and Eq. (\ref{condition}) becomes a condition on $m$. If we substitute the parameters used in Fig. \ref{horizons1.fig} we get that $m$ should be larger than a couple of hundreds, which does not seem to be experimentally realistic. However by choosing smaller values of $\beta_0$ and $w_0$, say by a factor of two or so, we can readily gain an order of magnitude or more. It means that we can arrive at the condition $m \sim 10$  which is challenging but should be feasible considering the state of the art of vortex beam generation techniques. Then instead of the exponentially small signal (\ref{spectraldensity-b}) we may expect much stronger signal (\ref{spectraldensity-a})
\beq
J \approx \frac{\lambda_r}{\pi^2 f}
\eeq
A possible shielding of the Hawking radiation and restriction on the possible $m$  values are discussed in Appendix C. These calculations also indicate that $|T|^2 \sim 1$.

\section{Summary and Conclusions}\label{conclusions}

In this work, we have used a coherent Laguerre-Gaussian beam propagating in a defocusing Kerr nonlinear medium as a model system for observation of the analogue Hawking radiation. Our approach is based on the hydrodynamic formulation of the propagation of light in a nonlinear medium [see Eqs.  \eqref{eqs2}] and it is therefore formally analogous to the dynamics of a compressible inviscid liquid. Compared to other models dealing with vortices, our model has the advantage of admitting nonzero radial flow by simply placing a diverging lens in font of the nonlinear medium itself. The diverging lens, in fact, induces a nonzero phase front curvature proportional to $r/f$ [see Eq. \eqref{eq11b}], which allows for a control of the radial flow and allows the formation of an event horizon in our model. The geometry induced by this vortex background gives rise to the situation depicted in Fig. \ref{flowscheme.fig}, where a white ($h_-$) and black ($h_+$) event horizon appear, together with two corresponding ergoregions ($e_-$ and $e_+$, respectively).

Considering the Hawking radiation from the (black) event horizon, we have shown that the vorticity of the background and of the field fluctuations compete to create a resonant amplification of the emitted Hawking radiation [see Eq. \eqref{radiation}]. Accounting for the leading nonadiabatic (i.e., slowly $z$-dependent) corrections result in important new features of the fluctuations, such as their enhancement or suppression in the vicinity of the horizon, whose magnitude can be controlled experimentally by varying the focal length of the diverging lens in the proposed experimental setup.

The most interesting new feature is the prediction of a resonance condition which may significantly amplify the otherwise extremely weak Hawking radiation in the relevant spectral interval. The same resonance condition controls the onset of SR, with total reflection taking place exactly at the resonance. Our estimates show that satisfying the conditions for experimental observation of the resonance, while challenging, is nevertheless a feasible task.
\section*{Acknowledgements}

The authors are indebted to the German -- Israeli Foundation. M. O. is thankful to the Raymond and Beverly Sackler Faculty of Exact Sciences at Tel-Aviv University for kindly hosting him while working on this project. The authors thank the Deutsche Forschungsgemeinschaft (grant BL 574/13-1) for financial support. Discussions with N. Pavloff and B. Reznik are highly appreciated.

\appendix

\section{Introduction of Radial Flow in a Laguerre-Gaussian Beam}\label{radial}
In this Appendix we explicitly derive Eqs. \eqref{eqs11} by using standard Gaussian optics. Let us consider the ABCD matrix describing the propagation of a Laguerre-Gaussian beam whose waist ($z=0$) coincides with the position of a defocusing lens of focal length $-f$. The input plane of the nonlinear medium is a short distance $z$ behind the lens. Thus
\beq
\left(
\begin{array}{cc}
A & B\\
C & D
\end{array}
\right)=\left(
\begin{array}{cc}
1 & z\\
0 & 1
\end{array}
\right)\left(
\begin{array}{cc}
1 & 0\\
1/f & 1
\end{array}
\right).
\eeq
We then use the self-similarity of the Gaussian $q$-parameter, i.e.
\beq
q'(z)=\frac{Aq(0)+B}{Cq(0)+D},
\eeq
to calculate the beam parameters. Recalling that
\beq
\frac{1}{q(z)}=\frac{1}{R(z)}-i\frac{2}{\beta w^2(z)},
\eeq
we can calculate the beam waist and the beam curvature at a distance $z$ from the lens, assuming that this distance is small compared to the Rayleigh range $z_R$ of the beam itself. Note that in the paraxial approximation this condition is easily satisfied, as the typical value of $z_R$ for a collimated beam of a few mm diameter is on the order of several meters. We therefore have, in the limit of small $z$
\barr
w^2(z)&\simeq&\frac{fw_0^2}{f-2z},\\
\frac{1}{R(z)}&\simeq&\frac{1}{f}.
\earr
If we now substitute these equations in the expressions for $\rho$ and $\vett{v}$ given by Eqs. \eqref{density-b}, and expand those in the limit of small $z$, we obtain the expressions used in Section \ref{III}.
\vspace{0.5cm}
\section{Superradiance}\label{AppC}

In this Appendix, we investigate the occurrence of SR in the scattering of electromagnetic fluctuations from the ergoregion. This effect has been discussed in GR \cite{S73,U74} and in several analogue systems \cite{BM03,RPLW15,BCP15}, including optical ones \cite{M08,MCO09}. Here, we shortly reproduce these results as applied to our system. The metric defined in Eq. \eqref{interval} allows us to write the differential equation in tortoise coordinates for the field fluctuation $\xi=\bar\xi\, e^{i\nu z- i m \phi}$  as follows:
\begin{eqnarray}\label{diffeq}
&&\frac{D}{r}\partial_r r D\partial_r\overline{\xi}
+ \widetilde{\nu}_m(r)^2\overline{\xi}\nonumber\\
&-& \frac{m^2D}{r^2} \overline{\xi}
- \frac{i \widetilde{\nu}_m(r)}{s}\left[\partial_z \ln(s^2D)\right] \overline{\xi}= 0,
\end{eqnarray}
where $D=(s^2-v_r^2)/s^2$. The last term in Eq. (\ref{diffeq}) accounts for the nonadiabatic evolution due to the $z$-dependence of the vortex profile. It also causes a weak dependence of $\overline{\xi}$ on $z$. However, this non adiabatic correction can be shown to be of order $1/(\nu z_R)\ll 1$ and can therefore be neglected in our analysis. Moreover, the divergence at $r \to r_h$ and $D \to 0$ occurs in a narrow region where a regularization procedure (such as the one outlined  in Section \ref{IV}) should be applied. Introducing the new coordinate $dr_*= D^{-1} dr$ and the new function $\psi = r^{1/2}\overline{\xi}$, Eq. (\ref{diffeq}) becomes
\begin{equation}\label{diffeq-a}
\partial_{r_*}^2 \psi + V(r(r_*)) \psi = 0,
\end{equation}
where $V(r(r_*))$ is the effective potential, whose explicit expression reads.
\beq\label{barrier}
V(r(r_*)) = \widetilde{\nu}_m(r)^2 - \frac{m^2D}{r^2(r_*)} +  \frac{1}{4}\frac{d}{dr} \left[\frac{ D^2}{ r(r_*)}\right].
\eeq
A closer inspection of the above equation, reveals that the effective potential has two $r_*$ independent asymptotes, namely
\bseq
\begin{align}
V(r) &\to \frac{1}{s^2}\left(\nu
 - \frac{m n}{\beta_0 r_h^2}\right)^2, & r\to r_h \\
 V(r) &\to \frac{\nu^2}{s^2}, & r\to\infty
\end{align}
\eseq
Corresponding to these asymptotic limits for the potential $V(r(r_*))$, the wavefunction $\psi(r_*)$ can be written as follows:
\bseq
\begin{align}
  \psi(r_*) &= e^{-i \nu r_*} + R e^{i \nu r_*}, &   r \to \infty, \label{inR}\\
  \psi(r_*) &= T \exp{\left[- i \left(\nu- \frac{m n}{\beta_0r_h^2}\right) r_*\right]} &  r \to r_h, \label{T}
\end{align}
\eseq
where $R$ and $T$ are reflection and transmission amplitudes, respectively. To calculate the relation existing between $R$ and $T$, we observe that since Eq. \eqref{diffeq-a} does not contain first derivatives, then, according to Abel's theorem, its Wronskian is constant \cite{byron}. Therefore, by equating the Wronskians calculated for the two above limits  by means of the functions (\ref{inR}) and (\ref{T}) and their complex conjugate, we get the following relation:
\begin{equation}\label{SR-a}
1 -|R|^2 =  \frac{|T|^2}{\nu}\left(\nu
- \frac{m n}{\beta_0 r_h^2}\right).
\end{equation}
Note that the same condition can also be obtained by balancing the ingoing and outgoing currents from the ergoregion. From the above equations one can readily see that the transmission coefficient
\beq
{\cal T} =  \frac{|T|^2}{\nu}\left(\nu
- \frac{m n}{\beta_0 r_h^2}\right)
\eeq
 may become negative. In this case, the reflection coefficient  ${\cal R}=|R|^2$ becomes larger than one,  meaning that the reflected wave is stronger than the incident wave, i.e. that the former is amplified. This is SR and in our system it takes place when the vorticity $n$ of the background vortex and the orbital angular momentum $m$ of the incident wave have the same sense and satisfy the condition
\begin{equation}\label{SR}
  \nu < \frac{m n}{\beta_0 r_h^2}.
\end{equation}

It is known that SR  in a real rotating black hole is dominated by the low $m$ components \cite{U74}. A way to see it is to consider a classically forbidden region with negative kinetic energy of the emitted particle \cite{M16,C68} which can shield the Hawking radiation or SR from escaping to the distant observer. In order to verify the occurrence of this scenario in our case, we can apply the rather complicated procedure developed in Refs. \cite{M16,C68}. However, the same result can be also obtained in a much simpler, although less rigorous, way, by using the eikonal approximation. To do this, we start from the Lagrangian corresponding to the metric \eqref{interval}, namely $\mathcal{L}= (1/2)g_{\mu\nu} \dot {x}^{\mu} \dot{x}^{\nu}$. From the Lagrangian, we can derive the canonical momentum $p_r=\dot{r}/D$. Then, by substituting $p_r$ for the momentum operator $-i\partial_r$ in Eq. \eqref{diffeq} and neglecting the small nonadiabatic terms, we get the following result:
\beq
\dot{r}^2=\left(\nu-\frac{mn}{\beta_0r^2}\right)^2- \left(\frac{D}{r^2}\right)\frac{m^2}{\beta_0^2r^2}.
\eeq
If we now expand the above expresison near the resonance $\nu_r=(mn)/(\beta_0r_h^2)$, we get
\beq\label{eqB9}
\dot{r}^2=\left(\delta\nu+2\nu_r\frac{x}{r_h}\right)^2+\varrho x\frac{\nu_r^2}{n^2},
\eeq
where $\delta\nu=\nu-\nu_r$ and $x=r-r_h$. Notice that the r.h.s. of the above equation becomes negative for $x_1<x<x_2<0$, where $x_{1,2}$ are the roots of the r.h.s. of Eq. \eqref{eqB9}, whose explicit expression reads
\beq
\frac{x_{1,2}}{r_h}=-\frac{1}{2}\left(\frac{\delta\nu}{\nu_r}+\frac{\varrho r_h}{4n^2}\right)\pm\frac{1}{2}\sqrt{\frac{\varrho r_h \delta\nu}{2n^2\nu_r}+\frac{\varrho^2 r_h^2}{16n^2}}.
\eeq
This area is classically forbidden and may shield the Hawking radiation \cite{M16}. This effect, however, can be offset by tunneling, whose probability is proportional to $e^{-2\lambda}$, where
\beq
\lambda=\left|\int_{x_1}^{x_2}\,dx\,p_r(x)\right|=\frac{\pi m}{4n\beta_0 r_h}+\mathcal{O}(\delta\nu).
\eeq
For the sake of simplicity, the above value of $\lambda$ has been obtained assuming perfect resonance (i.e., $\delta\nu=0$). Since we are interested in the situation when $\lambda<1$, i.e., when the barrier becomes transparent, from the above equation we may derive a condition on $m$ to be sufficiently small for that to occur, i.e.,
\beq
m<\frac{2\sqrt{2}\beta_0w_0n(n +1)^{1/2}}{\pi}\simeq 10^4n^{3/2},
\eeq
where we make use of (\ref{horizon-f}). If we now substitute typical values for the parameters, as the one used throughout this manuscript, we notice that the above constraint on $m$ is hardly restrictive and therefore rather large values of $m$ may be treated as "small".

\section{Nonadiabatic corrections}\label{nonadiabatic}

The coefficients in Eqs. (\ref{eqs4}) and (\ref{KG1}) derived in the main text depend on $z$. However, their derivation did not require any sort of adiabatic approximation. The necessity of applying an adiabatic approximation and investigating nonadiabatic corrections arise only when looking for solutions of these equations. It is the aim of this appendix to present this analysis, both for the case in which the quantum potential is neglected, and for the case in which it is taken into account. Furthermore, in the latter case we give the full expressions for the quantities appearing in Eqs. \eqref{eq36}, rather than only their leading order in $l_n^2$.
\subsection{Neglecting the quantum potential}
We start our analysis by considering Eq. \eqref{KG1}, which is obtained after neglecting the quantum potential. If we assume that the solutions to Eq. \eqref{KG1} can be sought in the form
\beq
\xi(r,z,\phi) = \overline{\xi}_{m,\nu}(x)e^{i\nu z-im\phi},
\eeq
where $x = r -r_h$ is the distance from the horizon, then Eq. \eqref{KG1} can be rewritten as follows:
\begin{widetext}
\beq\label{KG2}
\left[2 s_h^2 \widetilde{\nu}_m^2 + \left(\partial_z v_r\right) \partial_x + i\widetilde{\nu}_m s_h v_r \partial_x + i\frac{1}{r} s_h\partial_x r \widetilde{\nu}_mv_r + \frac{1}{r}\partial_x r \left(v_r^2 - s^2\right)\partial_x- \frac{m^2}{r} \left(v_r^2 + v_\phi^2- s^2\right)\right] \overline{\xi}_{m,\nu} = 0,
\eeq
\end{widetext}
where $s=s(x,z)$ is the background sound velocity, $v_r=v_r(x,z)$ and $v_{\phi}=v_{\phi}(r_h)$ are the radial and the azimuthal velocities, respectively, and
\beq
\widetilde{\nu}_m =\frac{1}{s_h}\left[\nu - \frac{mv_\phi}{r_h}\right] =\frac{1}{s_h}\left[\nu - \frac{mn}{\beta_0 r_h^2}\right].
\eeq
We seek solutions of Eq. \eqref{KG2} close to the event horizon, in the form $\overline{\xi}_{m,\nu} = e^{- ikx}x^{\gamma_m}$.  For the purposes of our analysis we consider the quantity  $r_h k$ to be large and keep the terms not higher than $O(r_h^{-2})$. The terms proportional to $1/r_h$ reflect the curvature of the event horizon. In the case of a stationary background these terms could also be neglected, as their contribution is a higher-order perturbation to the background solution. In our case, however, since the background is quasi-stationary, we keep these terms, as they play a non-negligible role in the determination of the correct quasi-stationary solution.

If we now substitute the ansatz $\overline{\xi}_{m,\nu} = e^{- ikx}x^{\gamma_m}$  into Eq. (\ref{KG2}) and collect terms with the same power of $x$ (with particular attention to $x^{-1}$), we obtain the following two values for $\gamma_m$:
\bseq
\begin{align}
\gamma_m^{(1)} &= 0, \\
\gamma_m^{(2)} &= -\frac{\alpha_r - \alpha_s + 2i\widetilde{\nu}_m }{s_h(2\alpha_r + \alpha_s)}.
\end{align}
\eseq
For the solution $\gamma_m^{(1)} =0$, we also get the following expression for the corresponding $k$ vector:
\beq
k_m  = \frac{i \widetilde{\nu}_m^2 + \widetilde{\nu}_m \alpha_r+\widetilde{\nu}_m/r_h}{2i\widetilde{\nu}_m  +  \alpha_r + 2\alpha_s }.
\eeq

\subsection{Accounting for the quantum potential}
We now turn our attention to the full solution of  Eqs. (\ref{eqs4}) in the case in which the quantum potential cannot be neglected. We apply the expansion outlined in Sect. \ref{IV} and then carry out the Fourier transform of $\chi(x,\phi,z)$, namely
\beq
\chi(x,\phi,z) = \sum_m \int  d\nu \int dk\, \chi_{k,m,\nu}\, e^{i(\nu z - m\phi -k z)}.
\eeq%
Substituting this into Eqs. \eqref{eqs4}, we obtain Eqs. \eqref{eqs4-1}. We can now solve the first of Eqs. \eqref{eqs4-1} with respect to
$\xi_{k,m,\nu}$ and substitute the result into the second of Eqs. \eqref{eqs4-1}, thus obtaining

\barr\label{solution-1}
\partial_k \ln\chi_{k,m,\nu} &=&
\frac{i\left[\widetilde{\nu}_m^2 - 2 \widetilde{\nu}_m k
   -  ik( \alpha_r - \alpha_s + 1/r_h) - \frac{m^2}{ r_h^2}\right]}{ k [2 \widetilde{\nu}_m \alpha_r - k (\alpha_s +2\alpha_r) ]}\nonumber\\
   &-&\frac{il_n^2}{2}\frac{ \left[( - \alpha_s + 1/r_h) ik  + k^2 + \frac{m^2}{ r_h^2}\right]^2 }{ k [2 \widetilde{\nu}_m \alpha_r - k  (\alpha_s +2\alpha_r) ]},
\earr
where $l_n^{-2}= 2\beta_0^2s_h^2$ is the nonlinearity length. If we now integrate the above equation, we can write the solution in the following form:
\begin{equation}\label{solution-2}
\chi_{m,\nu} =
e^{i(- m\phi+\nu z)}F(\tilde\nu,x),
\end{equation}
where $F(\tilde\nu,x)$ is given by Eq. \eqref{integral35} and reads
\beq
F(\tilde\nu,x)=\int_C\, dk\, k^{\gamma_1} \left(k - \frac{2 \widetilde{\nu}}{3}\right)^{\gamma_2} e^{\Lambda_{m,\nu}(k)-ikx}.
\eeq
The exact expressions for $\gamma_{1,2}$ and $\Lambda_{m,\nu}(k)$ are reported below, for the sake of completeness:
\bseq
\begin{align}
\gamma_1&= \frac {i\widetilde{\nu}_m }{2\alpha_r} - \frac {i {m}^{2}}{{2 r_h}^{2}\widetilde{\nu}_m\alpha_r}- \frac {i {l_n}^{2}{m}^{4}}{{4 r_h}^{4}\widetilde{\nu}_m\alpha_r},\\
\gamma_2&=\Bigg[\frac{\alpha_s-\alpha_r}{\varrho}-\frac{1}{r_h\varrho}- \frac{i\tilde\nu_m}{\varrho}\nonumber \\
&+\frac{2i\tilde\nu_m}{\varrho}+\frac{im^2}{2r_h^2 \tilde\nu_m\alpha_r}\Bigg]+l_n^2\Bigg[\frac{i\tilde\nu_m\alpha_r}{r_h^2\varrho^2}\nonumber\\
&+\frac{im^4}{4r_h^4\tilde\nu_m\alpha_r}+\frac{4i\tilde\nu_m^2\alpha_r^2}{r_h \varrho^3}+\frac{4\alpha_r^2\tilde\nu_m^2}{\varrho^2}\nonumber\\
&-i\tilde\nu_m\alpha_r- \frac{2\alpha_rm^2}{r_h^2\varrho}-\frac{4\tilde\nu_m\alpha_r^2}{r_h \varrho^2}+\frac{2\tilde\nu_m\alpha_r}{r_h\varrho}\nonumber\\
&+\frac{im^2}{r_h^3\varrho}+\frac{4i\tilde\nu_m^3\alpha_r^3}{\varrho^4}- \frac{4i\tilde\nu_m\alpha_r^3}{\varrho^2}+\frac{4i\tilde\nu_m\alpha_r^2}{\varrho} \nonumber\\
&+\frac{m^2}{r_h^2}-\frac{8\tilde\nu_m^2\alpha_r^3}{\varrho^3}+ \frac{2i\tilde\nu_m\alpha_rl^2m^2}{r_h^2\varrho^2}\Bigg],\\
\Lambda_{m,\nu}(k) &=\frac{k^3l_r^3}{3}\Bigg[\frac {i }{6}  + \frac {i}{ 2kr_h}  + \frac {i\widetilde{\nu} \alpha_r}{2 k\varrho} + \frac{\alpha_s}{2k}\nonumber\\
&+ \frac {2 i\alpha_r\nu}{ k^2r_h\varrho} + \frac {2\alpha_r\alpha_s\widetilde{\nu}}{k^2\varrho} + \frac {i}{2k^2 r_h^{2}}  + \frac {i{m}^{2}}{k^2r_h^{2}}\nonumber\\
&+\frac {2i{\alpha_r}^{2}{\widetilde{\nu}}^{2}}{k^2{\varrho}^2} + \frac {\alpha_s}{k^2r_h} - \frac{i \alpha_s^2}{2k^2} \Bigg],
\end{align}
\eseq
with $l_r=l_n(\varrho l_n/3)^{-1/3}$ being the regularisation length \cite{FS11} for the $z$-nonstationary flow.
\subsection{Normalization of the eigenfunctions}
\label{norma}
The goal of this appendix is to derive the normalisation constant of the eigenfunctions $\xi$ and $\chi$, and show that it depends, in the case of a vortex background, on the sign of $\tilde\nu$. To do that we rely on the approach described in Ref. \cite{FS11}, and define the two-component field
\beq
\psi=\left(\begin{array}{c}
\frac{\chi}{\sqrt{2}}\\
\\
\sqrt{2}\xi
\end{array}\right).
\eeq
Using standard field theory, we express the density and current associated to the field $\psi$ as follows:
\bseq\label{currDens}
\begin{align}
\rho^c&=& if_0^2(\xi^*\chi - \chi^* \xi),& \label{density-a} \\
{\bf j}^c &=& \rho^c{\bf v} - i \frac{f_0^2}{\beta_0}\left(\frac{1}{4}\chi^*\nabla_\perp\chi \right.&\left.+ \xi^*\nabla_\perp\xi - c.c.\right).\label{current}
\end{align}
\eseq

We then use the relation $\chi = (1/gf_9^2) \widehat{\mathcal D}\xi$, which holds at $|x|> l_n$, write Eqs. \eqref{currDens} in polar coordinates, and make the transformation Eqs. \eqref{transf},  obtaining:
\bseq\label{currents}
\begin{align}
\tilde\rho^c&= i\frac{s^2}{s^2 - v_r^2}\left[\xi^* \left(\overset{\leftrightarrow}{\partial_z}+\frac{v_{\phi}}{r}\overset{ \leftrightarrow}{\partial_{\phi}}\right)\xi\right],\label{density-c}\\
\tilde{j}_r^c &=i(v_r^2-s^2)\xi^*\overset{\leftrightarrow}{\partial_r}\xi,\\
\tilde j_{\phi}^c&=\xi^*\left[\frac{iv_{\phi}s^2}{s^2-v_r^2} \overset{\leftrightarrow}{\partial_z}+ \frac{s^2(v_0^2-s^2)}{r(s^2-v_r^2)}\overset{\leftrightarrow}{\partial_{\phi}} \right]\xi,
\end{align}
\eseq
where the derivative operator $\overset{\leftrightarrow}{\partial_x}$ is defined as follows:
\beq
\psi^*\overset{\leftrightarrow}{\partial_x}\psi=\psi^*\partial_x\psi-\psi\partial_x\psi^*.
\eeq
We now use Eq. \eqref{density-c} to define the scalar product, and consequently the norm of the eigenfunctions:
\beq
\{\psi_k,\psi_k\}=\int d\tilde r\tilde\rho_k^c,
\eeq
where the subscript $k$ indicates the type of solution we are considering, namely regular (r), evanescent (e) or singular (s). We are particularly interested in the behavior of the singular solution close to the event horizon. Following the procedure described in Refs. \cite{DR76,DL08}, we write
\beq
N(\{\xi_{m,1s},\xi_{m,1s}\} + \{\xi_{m,2s},\xi_{m,2s}\}e^{\frac{2\pi}{s_h\varrho}\widetilde{\nu}}_m)= - \mbox{sign}(\widetilde{\nu}_m).
\eeq
Using the scalar product defined above, one can readily verify that
\beq
\{\xi_{m,1s},\xi_{m,1s}\} = \mbox{sign}(\widetilde{\nu}_m),
\eeq
and
\beq
\{\xi_{m,2s},\xi_{m,2s}\} = - \mbox{sign}(\widetilde{\nu}_m).
\eeq
Hence, the normalisation constant $N$, which serves as a spectral density of the Hawking radiation, is:
\beq\label{resultOUR}
N = (e^{\frac{2\pi\widetilde{\nu}_m}{s_h\varrho}} - 1)^{-1}.
\eeq

It is worth mentioning, that a similar result has been obtained in the context of GR as well, see Refs. \cite{H75,C98,U10,MS06,ANOP10}.

\end{document}